# A RE-EXAMINATION OF REAL INTEREST PARITY IN CEECs USING OLD AND NEW GENERATIONS OF PANEL UNIT ROOT TESTS

ClaudiuTiberiuAlbulescu[a,b*], Dominique Pépin[b], AviralKumarTiwari[c]

[a] Management Department, Politehnica University Timisoara, 2, P-ta. Victoriei, 300006 Timisoara, Romania
[b] CRIEF, University of Poitiers, 2 rue Jean Carbonnier, 86022 Poitiers, France
[c] Faculty of Applied Economics, ICFAI University Tripura, West Tripura, 799210 Pin, India

## ABSTRACT

This study applies old and new generations of panel unit root tests to test the validity of long-run real interest rate parity (RIP) hypothesis for ten Central and Eastern European Countries (CEECs) with respect to the Euro area and an average of the CEECs' real interest rates, respectively.When the panel unit root tests are carried out with respect to the Euro area rate, we confirm the results of previous studies which support the RIP hypothesis. Nevertheless, when the test is performed using the average of the CEECs' rate, our results are mitigated, revealing that the hypothesis of CEECs' interest rates convergence cannot be taken for granted. From a robustness analysis perspective, our findings indicate that the RIP hypothesis for CEECsshould be considered with cautions, being sensitive to the benchmark.

[*]Correspondence: ClaudiuTiberiuAlbulescu, PolitehnicaUniversity Timisoara, P-ta. Victoriei, No. 2, 300006, Timisoara, Romania. Tel: 0040-743-089759. Fax: 0040-256-403021. E-mail: claudiu.albulescu@upt.ro,claudiual@yahoo.com.

I. INTRODUCTION

Since the collapse of the Gold-Exchange Standard, trade controls and barriers to international portfolios investments were progressively raised, triggering a considerable increase of the flexibility and substitutability of financial assets at international level during the last four decades. The generated financial globalization is supposed to lead to the elimination of worldwide arbitrage opportunities out of the Gold-Exchange Standard, the real interest rates observed in different countries being thus in the position of adjusting themselves based on the same global real exchange rate. This equality of the real interest rates is known as the real interest rates parity (RIP).

Yet, the RIP resulting from the national markets' interdependence is not a simple financial integration indicator. It has been shown that the confirmation of RIP depends on the extent of the uncovered interest parity (UIP), the relative purchasing power parity (RPPP) and the Fisher equation in domestic and foreign countries. Hence, confirmation of RIP encompasses elements of both real and financial market integration and, as such, it can be viewed as a more general indicator of integration or convergence (Holmes, 2002). From a policy perspective, it is desirable that a country joining a monetary union sees its realinterest rate adjusting itself to those of its counterparties. In this context, the accession of a country to a monetary union entails minimal costs.On the contrary, if the condition related to the equality of the real interest rates is violated, countries adopting the same currency will experience asymmetries in their responses to monetary policy shocks.The validity of the RIP is thus subject to debates of considerable importance for countries having abandoned their monetary sovereignty or which intend to do so, and, from this point of view, they are of great interest for the CEECs,members of the European Union (EU), because most of these countries have still preserved their monetary sovereignty.

Numerous researches on RIP were carried out for developed countries. The first studies (Mishkin, 1984, Cumby and Obstfeld, 1984, Mark, 1985 and Cumby and Mishkin, 1986) found that the proofs in favor of the RIP were quite limited, and this was for sure due to the short data sets used for the tests, and to the adopted econometric methodology, ignoring the non-stationarity of the interest rate. Other studies considering the non-stationarity of the interest rate(Meese and Rogoff, 1988, Edison and Pauls, 1993, Goldberg *et al.* (2003) and Pipatchaipoom and Norrbin, 2010) also reached inconclusive results, due to the use of short data samplesand to the lack of explanatory power of the standard unit roots tests in these conditions (Campbell and Perron, 1991). In order to solve this problem, it is possible to retain into the analysis a much longer time period (Lothian, 2002, Obstfeld and Taylor, 2002 and Sekioua, 2008), to perform nonlinear unit root tests or tests which take into account possible structural changes (Ferreira and León-Ledesma, 2007, Baharumshah*et al.*, 2009 and Camarero*et al.*, 2010), or to proceed to panel data analyses (Holmes, 2002 and Wu and Chen, 1998). The tests results provedclearly in favor of the RIP for the developed countries.

As theintegration of capital markets affecting the 'peripheral' regions of the worldstarted in the 1990s, the empirical analysis of the RIP validity was extended to cover these regions. In this line, Baharumshah*et al.* (2005, 2011), Liew and Ling (2008) and Holmes *et al.* (2011) found evidence of RIP for Asian countries. Camarero*et al.* (2010) and Ferreira and León-Ledesma (2007) showed the presence of the RIP in a sample of industrial and emerging economies. On the contrary, Singh and Banerjee (2006) found that real interest rates in the emerging markets show some convergence in the long run, but real interest rate equality does not hold. In respect of the CEECs, Baharumshah*et al.* (2013), Su et al. (2012), Sonora and Tica (2010), Cuestas and Harrison (2010), Arghyrou*et al.*(2009) and Holmes and Wang (2008a, 2008b) reached, based on diversified econometric techniques, the conclusion that the RIP is valid and that these countries should integrate the European Union: Arghyrou*et al.*(2009) and Sonora and Tica (2010) usedunivariate unit root tests allowing structural breaks,

Holmes and Wang (2008a) applied panel unit root tests, Holmes and Wang (2008b) employed an approach where unit roots tests are embedded within a Markov regime-switching framework, Cuestas and Harrison (2010) and Su *et al.* (2012) applied an univariate nonlinear unit root test, while Baharumshah *et al.* (2013) used a panel stationarity test that allows for multiple breaks. These papers obtained consistent results, but this is not surprising having in mind the fact that it is always the US and the European rates which are used as reference interest rate. Would the RIP remain valid if a real interest rate, representative for the CEEC, is used as benchmark? In our opinion, caution is required in front of these results because the RIP should be tested considering a European rate as reference interest rate, which resumes to testing the integration of CEECs into the EU, and, at the same time, it is also convenient to test the convergence of CEECs' interest rates between them. For this purpose, we propose to test the CEECs' RIP using a comparison with the average real interest rate of the selected CEECs, an approach which was never before investigated, as far as we know.

As it is well known that unit root tests lack explanatory power when it comes for short data samples, we use panel data analysis[1]. Moreover, in order to prove the robustness of these tests for panel data, otherwise questionable (Hurlin, 2010), we perform on each panel a set of tests, which can be split in two groups, namely, 'first generation unit root tests', that are based on a cross-sectional independence assumption and 'second generation unit root tests', that allow for cross-section dependence (see for example Baltagi, 2005, Breitung and Pesaran, 2008, and Hurlin, 2010). The first generation of panel unit root tests employed in this study include the MW test (Maddala and Wu, 1999), the Choi test (Choi, 2001), the LLC test (Levin *et al.*, 2002) and the IPS test (Im *et al.*, 2003). Part of these tests have been successfully used, despite the strong assumptions on which they rely, in the first empirical studies on the topic of the RIP for developed countries: the IPS test was carried out by Holmes (2002) et Wu and Chen (1998), and the MW test was performed by Wu and Chen (1998). It seems interesting to see if similar results could be obtained when analyzing the RIP in the case of the CEECs. Nonetheless, the hypothesis of the cross-sections interdependence is a strong assumption which, in case it is not confirmed, might cause the first generation tests to accept by mistake the hypothesis of a unit root (see for example Hurlin, 2010). It is thus important to resort to second generation tests. Besides, all unit root tests based on panel data and applied for the RIP starting with the middle of the years 2000s (Baharumshah *et al.*, 2005 and 2013, Camarero *et al.*, 2010, Liew and Ling, 2008 and Singh and Banerjee, 2006) relax the cross-sectional independence assumption. The second generation tests we use in this study are the MP test (Moon and Perron, 2004), the Pesaran test (Pesaran, 2007) and the Choi tests (Choi, 2006). Apart these second generation standard tests, ranging between the best known tests, we also propose the use of more recent second generation tests, as for example those of Lupi (2011), Hanck (2013) and Costantini and Lupi (2013), which are particularly well suited for the case of short term panels and which have never before been employed to test the validity of the RIP. Furthermore, these tests donot need balanced panel data sets, so that individual time series may come in different lengths and span different sample periods, which can be very useful in practice.

The rest of the paper is organized as follows. Section II describes the theoretical framework and the data used in the empirical analysis. Section III describes the econometric methodology used in this study, and Section IV presents the empirical results. Section V concludes.

---

[1]The other two empirical studies on the RIP for the CEECs using unit root tests in panel data are those of Holmes and Wang (2008a) and Baharumshah *et al.* (2013). However, none of these tests proceeds to the robustness check we propose by the use of the CEECs' average real interest rate as reference interest rate.

## II. DEFINITION OF THE REAL INTEREST PARITY AND DATA ANALYSIS

*II.1. Real Interest Parity*

The real interest parity condition is derived in a classical way, from the *ex-ante* relative purchasing power parity, the Fisher relation for each country, and the uncovered interest parity condition. In an open economy, the RIP relies on the equilibrium in the goods and services market on the one hand and asset markets on the other hand.

Consequently, the basic postulate for the relationship between domestic ($i$) and foreign ($i^*$) interest rates, shows that substitutable financial assets denominated in domestic and foreign currencies are related according to the UIP relationship, such as:

$$\Delta s^e_{t,t+1} = i_t - i_t^* - \gamma_t, \tag{1}$$

where $\Delta s^e_{t,t+1}$ is the one-period ahead expected change in the nominal exchange rate, measured as the domestic price of the foreign currency, and where $\gamma_t$ designates the interest rate differential explained by the exchange risk premium and other factors, such as transaction costs and national tax rates differential (Holmes, 2002; Sarno, 2005). Because the last factors have a reduced importance as compared to the first one[2], $\gamma_t$ can be simply considered as a risk premium[3]. Indeed, taking into account the risk premium is particularly relevant in the case of the CEECs (Sonora and Tica, 2010) or in the case of the emerging markets (Ferreira et León-Ledesma, 2007). Thus, Equation (1) is a general expression of the UIP, often simplified by the omission of the risk premium, that is $\gamma_t \equiv 0$.

The Fisher conditions for the domestic and foreign country are:

$$r_t^e = i_t - \Delta \pi_{t,t+1}, \tag{2}$$

and

$$r_t^{*e} = i_t^* - \Delta \pi_{t,t+1}^*, \tag{3}$$

where $r_t^e$ and $r_t^{*e}$ are the real interest rates in the domestic and foreign country respectively, while $\Delta \pi_{t,t+1}$ and $\Delta \pi_{t,t+1}^*$ are the one-period ahead expected inflation rates and $\pi$ is expressed as the natural logarithm of the price level (in our case, the consumer price index – CPI).

The *ex-ante* RPPP suggests that the exchange rate responds to offset spreads in expected inflation between countries. Otherwise said, the expected exchange rate depreciation should be equal to the expected inflation differential over the same period. Actually, the RPPP can be violated because of transaction costs and non-traded goods. We can thus formulate it as an imperfect short term relationship:

$$\Delta s^e_{t,t+1} = \Delta \pi_{t,t+1} - \Delta \pi_{t,t+1}^* + \varepsilon_t, \tag{4}$$

where $\varepsilon_t$ is the short term deviation from the RPPP. If we consider $\varepsilon_t \equiv 0$, the RPPP is verified both in the short and long runs.

Consequently, combining Equations (1)–(4), we reach the following result with respect to the real interest rates differential:

$$r_t^e - r_t^{*e} = \gamma_t + \varepsilon_t \tag{5}$$

Equation (5) has to be considered as a generalized expression of the RIP. Assuming that the risk premium is equal to zero, $\gamma_t \equiv 0$, and that the RPPP is always verified, $\varepsilon_t = 0$, we obtain the strong version of the RIP which stipulates a perfect equality between the real interest rates:

$$r_t^e - r_t^{*e} = 0 \tag{6}$$

where $r_t^e - r_t^{*e}$ represents the real interest rate differential (RIRD), equal to zero according to the strong version of the RIP. However, the existence of a risk premium and shortrun deviations from the RPPP prevents the RIRD from being constant at every point. If we assume that the risk premium $\gamma_t$ is stationary and that the RPPP describe a long run equilibrium relationship, the generalized expression of the RIP in Equation (5) supposes that the RIRD is a stationary variable:

$$r_t^e - r_t^{*e} \sim I(0) \qquad (7)$$

---

[2] Al-Awad and Grennes (2002) show that observed transactions for a group of 10 countries tend to decrease over time and are too small to account for differences among real interest rates.

[3] For simplifying the notations, $\gamma_t$ is the risk premium without referring to a specific pair of countries. Nevertheless, in general, this risk premium varies between countries.

Supposing that the domestic and foreign interest rates are integrated of order 1, $r_t^e \sim I(1)$ and $r_t^{*e} \sim I(1)$, the stationarity of their difference shows that the RIP is verified in the long run, that is, the RIRD is mean reverting. Thus, in order to test the validity of the longrun RIP equality[4], we must see whether the RIRD is stationary, as in Equation (7).

Because data on expected inflation is not readily available, unit root results are sensitive to how the inflation expectations and the real rate interest are computed (Pipatchaipoom and Norrbin, 2010). In order to assess the sensitivity of our results to the way we compute the real interest rates, we follow Cuestas and Harrison (2010) and Sonora and Tica (2010) and use *ex-ante* (rational) and *ex-post* (fitted) inflation expectations. The former implies that we obtain expected values for future inflation. In this respect, we suppose that agents use previous inflation rates $\pi_t$ and $\pi_t^*$ to form their expectations of future inflation rates, that is $\pi_{t,t+1} = \pi_t$ and $\pi_{t,t+1}^* = \pi_t^*$. The latter assumes perfect forecasting skills which mean that inflation expectations $\pi_{t,t+1}$ and $\pi_{t,t+1}^*$ are equal to the achievedfuture inflation rates $\pi_{t+1}$ and $\pi_{t+1}^*$[5].

The last technical detail of our methodology regarding the unit root tests of the RIRD, resides in the use of two different benchmarks. First, in the line of the previous RIP analyses for the CEECs, we consider an international interest rate, namely the Euro area rate, as reference. We thus test the hypothesis of CEECs' integration into the EU. Second, we establish as benchmark the average level of indicators for the retained CEECs. Consequently, we test the hypothesis related to the homogeneity of CEECs' real interest rates and to the convergence towards a representative rate for these countries. Mathematically, the CEECs' real interest rates cannot be globally equal to the Euro area rate, without being equal amongst them, which can thus be considered as a prerequisite for their European integration.

*II.1. Data*

The monthly data are obtained from the IMF's International Financial Statistics database, and cover the period from January 2000 to April 2012. The inflation rate is computed based on the CPI, while for the nominal interest rate we have chosen the money market rate. The use of a short term rate allows to minimize,but not to completely eliminate, the exchange rate influence, which is an important risk element in the case of the long term rates(Ferreira and Leon-Ledesma, 2007). After a preliminary data analysis, in order to avoid the broken panel problem, only 10 CEECs were retained in our sample (Bulgaria, Cyprus, Czech Republic, Estonia, Lithuania, Latvia, Poland, Romania, Slovakia, Slovenia). All these countries are nowadays members of the EU and five of them have also joined the Euro area(Cyprus, Estonia, Latvia, Slovakia and Slovenia). The descriptive statistics for the RIRD, for all the analyzed cases (*ex-ante*, *ex-post*, Euro area and the average of the group as benchmark) are presented in Table 1.

---

[4] The hypothesis of the RIP validity in the long run corresponds to the associated hypotheses related to the risk premium $\gamma_t$ stationarity, with short term tendencies fluctuating around its long term value $\gamma$, and to an adjusted version of the RPPP, where the long term deviations from the RPPP equal a constant term $\varepsilon$, which is not necessarily zero.

[5] Testing the stationarity of the RIRD calculated based on *ex-post* inflation expectations, can be considered as a robustness check for the results obtained employing the classic *ex-ante* relative purchasing power parity condition.

TABLE 1
*Summary statistics for the RIRD*

|       | *ex-ante* (Euro area) | | | | *ex-ante* (average) | | | | *ex-post* (Euro area) | | | | *ex-post* (average) | | | |
|-------|------|------|------|------|------|------|------|------|------|------|------|------|------|------|------|------|
|       | Min | Max | Mean | SD | Min | Max | Mean | SD | Min | Max | Mean | SD | Min | Max | Mean | SD |
| BU    | -5.35 | 3.46 | -0.97 | 1.33 | -10.3 | 1.03 | -2.76 | 2.36 | -5.44 | 4.03 | -0.98 | 1.35 | -10.1 | 1.72 | -2.76 | 2.37 |
| CY    | -4.84 | 2.72 | -0.14 | 1.60 | -6.25 | 0.75 | -1.89 | 1.72 | -4.72 | 3.20 | -0.15 | 1.61 | -6.21 | 0.84 | -1.91 | 1.73 |
| CZ    | -1.71 | 3.84 | 0.25 | 0.93 | -5.59 | 1.35 | -1.50 | 1.54 | -1.94 | 3.64 | 0.24 | 0.91 | -5.56 | 1.49 | -1.50 | 1.56 |
| ES    | -2.03 | 6.06 | 0.27 | 1.21 | -5.80 | 4.41 | -1.51 | 1.92 | -1.98 | 5.73 | 0.29 | 1.25 | -5.82 | 4.44 | -1.49 | 1.93 |
| LI    | -4.63 | 3.01 | -0.54 | 0.91 | -6.52 | 0.27 | -2.18 | 1.66 | -4.46 | 1.68 | -0.56 | 0.84 | -6.03 | 0.45 | -2.19 | 1.63 |
| LV    | -3.26 | 20.64 | -0.20 | 2.20 | -9.14 | 15.38 | -2.04 | 2.32 | -3.18 | 20.33 | -0.20 | 2.15 | -9.10 | 15.36 | -2.03 | 2.31 |
| PL    | -2.26 | 14.48 | 4.27 | 3.85 | -3.53 | 9.37 | 2.48 | 2.57 | -1.70 | 14.33 | 4.26 | 3.82 | -3.21 | 9.29 | 2.48 | 2.53 |
| RO    | 1.26 | 55.35 | 13.19 | 12.21 | 1.40 | 48.55 | 11.39 | 10.68 | 1.26 | 55.28 | 13.16 | 12.21 | -0.54 | 48.50 | 11.29 | 10.70 |
| SK    | -4.51 | 4.92 | 0.96 | 1.73 | -5.76 | 2.11 | -0.83 | 1.40 | -4.46 | 4.91 | 0.95 | 1.73 | -5.55 | 2.20 | -0.83 | 1.41 |
| SV    | -2.54 | 5.19 | 0.84 | 1.14 | -4.67 | 2.47 | -0.93 | 1.48 | -2.62 | 5.16 | 0.83 | 1.13 | -4.63 | 2.59 | -0.95 | 1.48 |
| Panel | -5.35 | 55.3 | 1.87 | 5.95 | -10.3 | 48.55 | 0.10 | 5.64 | -5.44 | 55.28 | 1.87 | 5.94 | -10.1 | 48.50 | 0.10 | 5.63 |

*Note:* BU (Bulgaria), CY (Cyprus), CZ (Czech Republic), ES (Estonia), LI (Lithuania), LV (Latvia), PL (Poland), RO (Romania), SK (Slovakia), SV (Slovenia).

We first notice that, in all the cases, Romania is less integrated in the selected group. Its higher inflation during the 1990s forced the monetary authorities to practice high interest rate, with implications in 2000s also. Second, we observe that no matter the considered benchmark (Euro area or CEECs' average), the descriptive statistics are not sensitive to the way of measuring inflation expectations. The use of *ex-ante* or *ex-post* inflation expectations proved no influence on the results. Third, the choice of the reference interest rate has an important impact on the RIRD's computation. It seems that seven out of the ten considered countries present real interest rates much more divergent with the group average as compared to the Euro area interest rate. More precisely, only in the case of Poland, Romania and Slovakia the real interest rate is closer to the average of the group. This evidence indicates that the results of the unit root tests can be sensitive to the benchmark.

## III. METHODOLOGY

We test the RIP hypothesis for the CEECs using different types of unit root tests on panel data. We resort to the following first generation unit root tests: the MW test (Maddala and Wu, 1999), the Choi test (Choi, 2001), the LLC test (Levin *et al.*, 2002) and the IPS test (Im *et al.*, 2003), that are all based on the assumption of independent cross-section units. We then proceed to the following second generation unit root tests: the MP test (Moon and Perron, 2004), the Pesaran test (Pesaran, 2007) and the Choi tests (Choi, 2006). Finally, we propose recent second generation tests, which have not been previously used for testing the validity of the RIP: Lupi (2011), Hanck (2013) and Costantini and Lupi (2013).

The basic model underlying these tests is:

$$\Delta RIRD_{i,t} = \alpha_i + \rho_i RIRD_{i,t-1} + \sum_{k=1}^{p_i} \beta_{i,k} \Delta RIRD_{i,t-k} + v_t \qquad (8)$$

for $i = 1, \ldots, N$ and $t = 1, \ldots, T$.

For all these tests (with the exception of the LCC test), the null hypothesis is defined as $H_0: \rho_i = 0$ for all $i = 1, \ldots, N$ and the alternative hypothesis is $H_1: \rho_i < 0$ for $i = 1, \ldots, N_1$ and $\rho_i = 0$ for $i = N_1, \ldots, N$ with $0 < N_1 \leq N$. The alternative hypothesis allows unit roots for some (but not all) of the countries. In the particular case of the LCC test, we simplify the model (8) with the additional assumptions: $\alpha_i = 0$ and $\rho_i = \rho$ for all $i = 1, \ldots, N$. The null hypothesis is then defined as $H_0: \rho = 0$ for all $i = 1, \ldots, N$ and the alternative hypothesis is $H_1: \rho < 0$ for $i = 1, \ldots, N_1$.

*III.1. First generation of panel stationarity tests*

The first test included in the first generation of unit root tests is that of Maddala and Wu (1999). According to the authors, if the test's statistics are continuous, the significance levels $\pi_i, (i = 1, 2, \ldots N)$ are independent and uniform (0,1) variables. The MW test is based on the combined significance levels (*p*-values or $P_{MW}$) of the individual unit root tests:

$$P_{MW} = -2 \sum_{i=1}^{N} \log \pi_i \tag{9}$$

where $-2 \sum_{i=1}^{N} \log \pi_i$ has a $\chi^2$ distribution with 2N degrees of freedom.

Based on the MW test, Choi (2001) suggests the following standardized statistic, which, under the cross-sectional independence assumption, converges to a standard normal distribution:

$$Z_{MW} = \frac{\sqrt{N}\{N^{-1} P_{MW} - E[-2 \log(\pi_i)]\}}{\sqrt{Var[-2 \log(\pi_i)]}} \tag{10}$$

The third test described in the first generation category is the test of Levin *et al.* (2002), called LLC (Levin, Lin, Chu) test. The LLC test employs the following adjusted *t*-statistic:

$$t_{\partial}^{*} = \frac{t_{\alpha} - (NT)\hat{S}_N \sigma_{\tilde{\varepsilon}}^{-2} \sigma_{\tilde{\alpha}} \mu_T^{*}}{\sigma_T^{*}} \tag{11}$$

where $\hat{S}_N$ is the average of individual ratios in the longrun to shortrun variance for the country *i*, $\sigma_{\tilde{\varepsilon}}$ is the standard deviation of the error term, $\sigma_{\tilde{\alpha}}$ is the standard deviation of the slope coefficients, $\sigma_T^{*}$ is the standard deviation adjustment, $\mu_T^{*}$ is the mean adjustment.

The last test that we retain in the first generation category is the test of Im *et al.* (2003), which employs a standardized *t_bar* statistic based on the movement of the Dickey-Fuller distribution:

$$Z_{t\_bar} = \frac{\sqrt{N}\{t\_bar - N^{-1} \sum_{i=1}^{N} E(t_{iT})\}}{\sqrt{N^{-1} \sum_{i=1}^{N} Var(t_{iT})}} \tag{12}$$

where $E(t_{iT})$ is the expected mean of $t_{iT}$ and $Var(t_{iT})$ is its variance.

*III.2. Second generation of panel stationarity tests*

The first tests included in the second generation of unit root tests are those of Moon and Perron (2004). The authors use a factor structure to model cross-sectional dependence, assuming that error terms are generated by common factors and idiosyncratic shocks. The MP tests consider thus the factors as nuisance parameters and suggest pooling de-factored data to construct a unit root test. The two modified *t*-statistics with standard normal distribution under the null hypothesis are:

$$t_a = \frac{T\sqrt{N}(\rho_{pool}^{+} - 1)}{\sqrt{2\gamma_e^4/w_e^4}} \xrightarrow[T,N \to \infty]{d} N(0,1) \tag{13}$$

$$t_b = T\sqrt{N}(\rho_{pool}^{+} - 1)\sqrt{\frac{1}{NT^2} trace(Z_{-1} Q_\wedge Z_{-1}')\frac{w_e^2}{\gamma_e^4}} \xrightarrow[T,N \to \infty]{d} N(0,1) \tag{14}$$

where $w_e^2$ denotes the cross-sectional average of the longrun variances of residuals *e*, $\gamma_e^4$ is the cross-sectional average of $w_e^4$. Moon and Perron (2004) propose feasible statistics $t_a^{*}$ and $t_b^{*}$ based on an estimator of the projection matrix and estimators of longrun variances $w_{ei}^2$.

Pesaran (2007) proposes a test where the augmented Dickey-Fuller (ADF) regressions are augmented with the cross-sectional average of the lagged levels and the first-differences of the individual time series. This way, the common factor is proxied by the cross-section mean of $y_{i,t}$ and its lagged values. The Pesaran test uses the cross-sectional ADF statistics (CADF), which are given below:

$$\Delta y_{i,t} = \alpha_i + \beta_i y_{i,t-1} + \gamma_i \bar{y}_{t-1} + \delta_i \Delta \bar{y}_i + \varepsilon_{i,t} \tag{15}$$

where $\alpha_i, \beta_i, \gamma_i, \delta_i$ are slope coefficients estimated from the ADF test for the country *i*, $\bar{y}_{t-1}$ is the mean of lagged levels, $\Delta \bar{y}_i$ is the mean of first-differences, $\varepsilon_{i,t}$ are the error terms.

In fact, Pesaran (2007) advances a modified IPS statistics based on the average of the individual CADF, which is denoted as a cross-sectional augmented IPS (CIPS):

$$CIPS = \frac{1}{N}\sum_{i=1}^{N} t_i(N,T) \qquad (16)$$

where $t_i(N,T)$ is the *t*-statistic of the OLS estimate for the equation $y_{it} = \alpha_i + y_{it}^0$ (see Moon and Perron, 2004).

Choi (2006) uses an error-component model to specify the cross-sectional correlations. The author suggests that cross-sectional correlations and deterministic components are eliminated by the GLS-based detrending (Elliott *et al.*, 1996) and the conventional cross-sectional demeaning for panel data. Afterwards, Choi (2006) employs a standard ADF *t*-statistic on the regression Equation (8) without intercept, where the corrected RIRD is substituted for the original one. Based on these individual standard ADF *t*-statistics, Choi (2006) proposes three Fisher's type tests, based on:

$$P_m = -\frac{1}{\sqrt{N}}\sum_{i=1}^{N}[\ln(p_i) + 1] \qquad (17)$$

$$Z = \frac{1}{\sqrt{N}}\sum_{i=1}^{N}\Phi^{-1}(p_i) \qquad (18)$$

$$L* = \frac{1}{\sqrt{\pi^2 N/3}}\sum_{i=1}^{N}\ln\left(\frac{p_i}{1-p_i}\right) \qquad (19)$$

where $p_i$ denotes the asymptotic *p*-values of the standard ADF *t*-statistics for the country $i$, $\Phi^{-1}(.)$ is the inverse normal cumulative distribution function and $P_m$, $Z$ and $L*$ are the first, the second and respectively the third Choi's tests. Under the null hypothesis, all these three Fisher's type statistics have a standard normal distribution.

### III.3. New proposed panel stationarity tests

The first generation of panel unit root tests (Maddala and Wu, 1999; Choi, 2001) uses simple *p*-values combination tests and assumes that the panel units are cross-sectionally independent. In particular Choi (2001) suggests that, under the null, the inverse normal combination test has the best overall performance:

$$Z = \frac{1}{\sqrt{N}}\sum_{i=1}^{N}\Phi^{-1}(p_i) \xrightarrow{d} N(0,1) \qquad (20)$$

However, this assumption is quite restrictive and is exploited for the first time by Hartung (1999). Afterwards, Demetrescu *et al.* (2006), propose a modification of the Choi's inverse-normal combination test that can be used when the $N_p$ values are not independent.

In particular, Hartung (1999) shows that if the probits $\Phi^{-1}(p_i)$ are correlated, with a common correlation $\partial$ which is practically unknown, then, under the null:

$$Z_H = \frac{1}{\sqrt{N(1+\partial(N-1))}}\sum_{i=1}^{N}\Phi^{-1}(p_i) \sim N(0,1) \qquad (21)$$

Demetrescu *et al.* (2006) propose a common practical implementation in their simulations, which relies on the works of Hartung (1999) and Choi (2001):

$$\hat{Z}_H = \frac{\sum_{i=1}^{N}\Phi^{-1}(p_i)}{\left\{N\left[1+\left(\hat{\partial}*+0.2\sqrt{\frac{2}{N+1}}(1-\hat{\partial}*)\right)(N-1)\right]\right\}^{\frac{1}{2}}} \qquad (22)$$

where $\hat{\partial}*$ is a consistent estimator of $\partial$, such that $\hat{\partial}* = max\left\{-\frac{1}{N-1}, \hat{\partial}\right\}$, with

$$\hat{\partial} = 1 - (N-1)^{-1}\sum_{i=1}^{N}(\Phi^{-1}(p_i) - N^{-1}\sum_{i=1}^{N}\Phi^{-1}(p_i))^2 \qquad (23)$$

A rather different viewpoint is advanced by Hanck (2013), who observes that the problem of panel unit root testing can be recast in terms of a multiple testing problem, using the intersection test presented in Simes (1986). He considers $p_{(i)}$ the ordered sequence of the $N_p$ values of each unit root test on each individual series. The proposed test is simple to compute, as for a pre-specified significance level $\alpha$, the null is rejected if $p_{(i)} \leq i\alpha/N$ for any $i = 1, \ldots, N$.

Other developments are put forward by Costantini and Lupi (2013), who consider panel covariate Dickey-Fuller tests as simple extensions, based on the *p*-value combination methods outlined above, of the CADF test advocated in Hansen (1995). Hansen (1995) proves that the unit root test statistic under the null is no longer distributed according to a Dickey-Fuller distribution, but is instead distributed according to a weighted sum of a Dickey-Fuller and a standard normal distribution, where the weights are functions of a nuisance parameter. The testing equation is however very similar to the ordinary ADF equation:

$$a(L)\Delta y_t = \delta y_{t-1} + b(L)\Delta x_{t-1} + e_t \tag{24}$$

Under the fulfillment of some regularity conditions, Hansen (1995) shows that, under the unit root null, the $t$ ratio for the coefficient in Equation (24) is such that:

$$\hat{t}(\delta) \xrightarrow{\omega} \rho \frac{\int_0^1 W dW}{\left(\int_0^1 W^2\right)^{\frac{1}{2}}} + (1-\rho^2)^{\frac{1}{2}} N(0,1) \tag{25}$$

where $W$ is a standard Wiener process and $N(0,1)$ is a standard normal independent of $W$.

Costantini and Lupi (2013) exploit this idea. The *p*-value combination suggested by the authors follows however Choi (2001), when no cross-dependence is detected, and Demetrescu *et al.* (2006) in the presence of cross-dependence. Briefly, Costantini and Lupi (2013) suggest using the average of the first difference of the other series in the panel, as the stationary covariate for each variable to be tested. This procedure aims at extracting an underlying non-stationary common factor among the observed series. In this case, the panel CADF test explicitly refers to cross-dependent time series.

The use of Hansen's CADF test instead of the conventional ADF test ensures that the panel test has better explanatory power properties. It is named pCADF test. Furthermore, in Costantini and Lupi (2013), contrary to Demetrescu *et al*. (2006), the Hartung's procedure for cross-correlation correction is applied only when the *p*-value of the cross-correlation test advocated by Pesaran (2004) is lower than a pre-specified threshold whose default value is set to 0.10. Therefore, in the present study we apply the recent proposed test to the panel covariate augmented Dickey-Fuller test, proposed by Costantini and Lupi (2013). In addition, we also apply an extension of the CADF tests suggested by Hanck (2013) and developed in Lupi (2011). Given its relation to the Simes' (1986) procedure, we label the latter test as sCADF.

## IV. EMPIRICAL RESULTS

*IV.1. RIRD based on ex-ante inflation expectations*

In the first step, we present the results of the unit root tests for the RIRD, computed based on *ex-ante* inflation expectations, more exactly assuming that $\pi_{t,t+1} = \pi_t$ and $\pi^*_{t,t+1} = \pi^*_t$.

The results for the first generation of panel unit root tests are provided in Table 2. The MW test uses combined significance levels, and rejects the null hypothesis of unit root at 1% significance level, no matter if we consider the Euro area or the average of the group as benchmark for the RIRD calculation. The Choi test provides findings that are similar to the MW test, proving the stationarity of the RIRD. Furthermore, the LLC test clearly indicates that the nonstationarity is a common feature of the RIRD in the selected countries, contrary to the previous tests. Indeed, at all significance levels, the test does not reject the null of nonstationarity. However, the LLC test is criticized for its assumption that $\rho$ is homogeneous. In addition, we should not forget that the alternative hypothesis of this particular test is different as compared to the other tests, because it does not admit the nonstationarity condition of the RIRD for any of the selected countries. As a result, it is possible to reject the nonstationarity hypothesis for the entire panel of countries, if the RIRD for a single country is nonstationary. Finally, the IPS test confirms the results of the first two tests, showing the absence of unit roots. Consequently, the results based on the first generation tests are robust, and validate the PPP theory.

TABLE 2

*First generation of panel unit root tests for ex-ante inflation expectations*

| Tests | Test statistic | | Critical values | | |
|---|---|---|---|---|---|
| | Euro area as benchmark | Average of the group as benchmark | 1 % | 5 % | 10 % |
| MW test (Maddala and Wu, 1999) | 59.711 | 53.731 | 37.566 | 31.410 | 28.412 |
| Choi test (Choi, 2001) | 6.2789 | 5.3334 | 2.3263 | 1.6449 | 1.2816 |
| LLC test (Levin *et al.*, 2002) | 1.5193 | -0.2977 | -2.3263 | -1.6449 | -1.2816 |
| IPS test (Im *et al.*, 2003) | -3.5257 | -2.8204 | -2.3263 | -1.6449 | -1.2816 |

*Note: In order to reject the null of unit root, the MW and Choi statistics must be above their critical values, while the LLC and IPS statistics have to be under their critical values.*

The first generation of panel unit root tests is criticized for assuming cross-sectional independence, assumption which is relaxed under the second generation of panel unit root tests. In this case, the entire battery of second generation tests shows that the RIRD is stationary, at all significance levels (Table 3). Moreover, these results are robust regarding the retained benchmark. Nevertheless, an exception appears for the Pesaran's CIPS test, which does not reject the null hypothesis of unit root for the case where the RIRD is estimated based on the group's average as benchmark.

TABLE 3

*Second generation of panel unit root tests for ex-ante inflation expectations*

| Tests | Test statistic | | Critical values | | |
|---|---|---|---|---|---|
| | Euro area as benchmark | Average of the group as benchmark | 1 % | 5 % | 10 % |
| First MP test (Moon and Perron, 2004) | -20.526 | -13.788 | -2.3263 | -1.6449 | -1.2816 |
| Second MP test (Moon and Perron, 2004) | -5.7142 | -5.1461 | -2.3263 | -1.6449 | -1.2816 |
| First Choi's test statistic (Choi, 2006) | 16.427 | 14.006 | 2.3263 | 1.6449 | 1.2816 |
| Second Choi's test statistic (Choi, 2006) | -8.5340 | -7.8596 | -2.3263 | -1.6449 | -1.2816 |
| Third Choi's test statistic (Choi, 2006) | -10.738 | -9.4108 | -2.3263 | -1.6449 | -1.2816 |
| Pesaran's CIPS test (Pesaran, 2007) | -3.1107 | 2.9493 | -2.5669 | -2.3310 | -2.2062 |

*Note: In order to reject the null of unit root, the first Choi statistic has to be above its critical value, while all the other statistics must be under their critical values.*

Finally, the results are mixed in the case of the proposed panel unit root tests (Table 4). When analyzing the results of the tests proposed by Costantini and Lupi (2013), we notice that, in general, all three methods reassert the results of the first and the second generation of tests, showing that the null hypothesis of unit root is rejected. However, the pCADF test rejects the null only if the Euro area is considered as benchmark for the RIRD estimation. In addition, the pCADF_PC test confirms the stationarity when we consider the average as benchmark, but only at 10% significance level. These results lack completely in robustness when we look to the tests proposed by Hanck (2013) and Lupi (2011). When considering the average as benchmark, we observe that all three tests do not reject the null, proving thus the presence of a unit root.

The last results are very interesting because the tests of Costantini and Lupi (2013), Lupi (2011) and Hanck (2013) bring forward an original version of the CEECs' integration. If, in general, the hypothesis of their integration with the Euro area is accepted when assuming the European rate as benchmark, the hypothesis of the interest rate convergence for these countries towards a common rate is afterwards rejected. Thus, the first conclusion must be considered with caution due to the fact that the alignment of the CEECs' real interest rates with the European rate cannot be achieved without a convergence between the CEECs.

TABLE 4

*Proposed panel unit root tests for ex-ante inflation expectations*

| Tests | | Euro area as benchmark | | Average of the group as benchmark | |
|---|---|---|---|---|---|
| | Constant model | Test statistic | *p*-value | Test statistic | *p*-value |
| pADF (Costantini and Lupi, 2013) | | -3.959905e+0 | 3.748975e-0 | -2.5849672 | 0.0048694 |
| pCADF (Costantini and Lupi, 2013) | | -4.078241e+0 | 2.268882e-0 | -0.9266757 | 0.1770475 |
| pCADF_PC (Costantini and Lupi, 2013) | | -2.00413297 | 0.02252791 | -1.6283338 | 0.0517270 |
| | Decision on H0 | 1% | 5% | 10% | 1% | 5% | 10% |
| Simes ADF (Hanck, 2013) | | FALSE | FALSE | FALSE | TRUE | TRUE | TRUE |
| SimespCADF (Lupi, 2011) | | FALSE | FALSE | FALSE | TRUE | TRUE | TRUE |
| Simes CADF_PC (Lupi, 2011) | | TRUE | TRUE | FALSE | TRUE | TRUE | TRUE |

*Notes: (1) pADF test is proposed by Costantini and Lupi (2013) based on Choi (2001) when no cross-dependence is detected. (2) pCADF test is proposed by Costantini and Lupi (2013) based on Demetrescu et al. (2006) in the presence of cross-dependence. (3) pCADF_PC test is the Panel Covariate Augmented DF test, proposed by Costantini and Lupi (2013), who assume that the panel is balanced and utilize the differenced first principal component of the N series as the stationary covariate. In the present case we use max.lag.y = 5, max.lag.x = 5. (4) Simes ADF test is proposed by Hanck (2013), based on Simes (1986). (5) SimespCADF test is an ADF-based test proposed by Lupi (2011), advancing over Hanck (2013). (6) SimespCADF_PC test is an ADF-based test proposed by Lupi (2011) advancing over Hanck (2013) whichemploys the differenced first principal component of the N series as the stationary covariate. (7) TRUE indicates that the test does not reject the null and FALSE shows that the null is rejected. (8) In each case the lag selection is based on AIC information criteria and we fix the maximum number of lags to 5.*

## IV.2. RIRD based on ex-post inflation expectations

In order to assess the robustness of the previous results, we proceed to new empirical estimations, relyingonthe *ex-post* inflation expectations, when calculating the RIRD.In this case we assume that the inflation expectations are equal to the realized inflation, that is $\pi_{t,t+1} = \pi_{t+1}$ and $\pi^*_{t,t+1} = \pi^*_{t+1}$. The results for the first generation tests are similar to those obtained based on *ex-ante* inflation expectations (Table 5). We notice that the MW test, the Choi test and the IPS test indicate the stationarity of the RIRD in both cases. The only test which provides opposite results is that of Levin *et al.* (2002).

TABLE 5

*First generation of panel unit root tests for ex-post inflation expectations*

| Tests | Test statistic | | *p*-values | | |
|---|---|---|---|---|---|
| | Euro area as benchmark | Average of the group as benchmark | 1 % | 5 % | 10 % |
| MW test (Maddala and Wu, 1999) | 60.771 | 51.619 | 37.566 | 31.410 | 28.412 |
| Choi test (Choi, 2001) | 6.4466 | 4.9995 | 2.3263 | 1.6449 | 1.2816 |
| LLC test (Levin *et al.*, 2002) | 1.9053 | 0.1537 | -2.3263 | -1.6449 | -1.2816 |
| IPS test (Im*et al.*, 2003) | -3.7265 | -2.9273 | -2.3263 | -1.6449 | -1.2816 |

*Note: In order to reject the null of unit root, the MW and Choi statistics must be above their critical values, while the LLC and IPS statistics have to be under their critical values.*

The results of the second generation tests are identical with the findings obtained in the previous case(Table 6). All the tests confirm the rejection of the null and show that the RIRD is stationary. We can also notice that the Pesaran's CIPS test validates the stationarity at 5% significance level, even if we consider the group average as benchmark for estimating the RIRD.

TABLE 6

*Second generation of panel unit root tests for ex-post inflation expectations*

| Tests | Test statistic | | *p*-values | | |
|---|---|---|---|---|---|
| | Euro area as benchmark | Average of the group as benchmark | 1 % | 5 % | 10 % |
| First MP test (Moon and Perron, 2004) | -17.082 | -13.450 | -2.3263 | -1.6449 | -1.2816 |
| Second MP test (Moon and Perron, 2004) | -5.1345 | -4.5609 | -2.3263 | -1.6449 | -1.2816 |
| First Choi's test statistic (Choi, 2006) | 16.315 | 13.268 | 2.3263 | 1.6449 | 1.2816 |
| Second Choi's test statistic (Choi, 2006) | -8.4363 | -7.5275 | -2.3263 | -1.6449 | -1.2816 |
| Third Choi's test statistic (Choi, 2006) | -10.661 | -8.9831 | -2.3263 | -1.6449 | -1.2816 |
| Pesaran's CIPS test (Pesaran, 2007) | -3.1329 | -2.3753 | -2.5669 | -2.3310 | -2.2062 |

*Note: In order to reject the null of unit root, the first Choi statistic has to be above its critical value, while all the other statistics must be under their critical values.*

The results provided by the new generation of panel unit root tests are mixed in this case also. On the one hand, the tests proposed by Costantini and Lupi (2013) advance contradictory results. If for the first category of analyses (Euro area as benchmark), all three tests indicate the stationarity of the RIRD, for the second category of tests (average of the group as benchmark), only the pADF test confirms the stationarity, while pCADF and pCADF_PC tests do not reject the null of unit root. On the other hand, we can also notice that the Simes-based tests provide contradictory results when changing the benchmark. However, in this situation, all three tests confirm the stationarity at 5% significance level for the first group of analyses, considering the Euro area as benchmark, while for the second group of analyses all three tests assert the presence of a unit root.

TABLE 7

*Proposed panel unit root tests for ex-post inflation expectations*

| Tests | Euro area as benchmark | | Average of the group as benchmark | |
|---|---|---|---|---|
| Constant model | Test statistic | *p*-value | Test statistic | *p*-value |
| pADF (Costantini and Lupi, 2013) | -3.7084505 | 0.0001042 | -2.3303113 | 0.0098948 |
| pCADF (Costantini and Lupi, 2013) | -2.1752106 | 0.0148071 | -0.8349678 | 0.2018679 |
| pCADF_PC (Costantini and Lupi, 2013) | -2.9547189 | 0.0015647 | -0.5041164 | 0.3070898 |
| Decision on H0 | 1%  5%  10% | | 1%  5%  10% | |
| Simes ADF (Hanck, 2013) | TRUE  FALSE  FALSE | | TRUE  TRUE  TRUE | |
| SimespCADF (Lupi, 2011) | TRUE  FALSE  FALSE | | TRUE  TRUE  TRUE | |
| Simes CADF_PC (Lupi, 2011) | TRUE  FALSE  FALSE | | TRUE  TRUE  TRUE | |

*Notes: (1) pADF test is proposed by Costantini and Lupi (2013) based on Choi (2001) when no cross-dependence is detected. (2) pCADF test is proposed by Costantini and Lupi (2013) based on Demetrescu et al. (2006) in the presence of cross-dependence. (3) pCADF_PC test is the Panel Covariate Augmented DF test, proposed by Costantini and Lupi (2013), who assume that the panel is balanced and utilize the differenced first principal component of the N series as the stationary covariate. In the present case we use max.lag.y = 5, max.lag.x = 5. (4) Simes ADF test is proposed by Hanck (2013), based on Simes (1986). (5) SimespCADF test is an ADF-based test proposed by Lupi (2011), advancing over Hanck (2013). (6) SimespCADF_PC test is an ADF-based test proposed by Lupi (2011) advancing over Hanck (2013) which employs the differenced first principal component of the N series as the stationary covariate. (7) TRUE indicates that the test does not reject the null and FALSE shows that the null is rejected. (8) In each case the lag selection is based on AIC information criteria and we fix the maximum number of lags to 5.*

All in all, we can state that the panel unit root tests confirm in general the outcomes of the previous studies on the RIP for the CEECs, when the European rate is considered as benchmark, and also the results of the studies carried out on panel data, namely Holmes and Wang (2008a) and Bararumshah*et al.* (2013). Our findings sustain thus the hypothesis that the CEECs' real interest rates are integrated with the European rate, and we obtain this result without resorting to structural breaks as it is the case in the analysis of Bararumshah*et al.* (2013).

However, the results of the panel unit root tests when considering the average interest rate as benchmark are much more mitigated. The first generation tests and the classical tests from the second generation prove the RIRD stationarity. On the contrary, the tests of Lupi (2011), Hanck (2013) and Costantini and Lupi (2013), well adapted for small panels, question the RIP hypothesis for the CEECs. According to their results, we cannot state that the CEECs' real interest rates are convergent towards a single common rate.

## V. CONCLUSIONS

The purpose of this study was to perform an empirically analysis of the RIP for ten CEECs, relying on panel unit root tests, considering *ex-ante* and *ex-post* inflation expectations, and successively using the Euro area and the group average interest rates as benchmarks. Our results underline the RIP's lack of sensitivity in respect of the way inflation expectations are defined (*ex-ante* or *ex-post*), but a moderate sensitivity regarding the selected benchmark. The RIP hypothesis is clearly accepted when considering the Euro area rate as benchmark, showing thus the CEECs' integration into the EU. However, the findings are much more mitigated when using the group average as reference. In this line, according to the tests of Lupi (2011), Hanck (2013) and Costantini and Lupi (2013), well suited for small panels, the RIRD stationarity hypothesis is rejected.

The fact that the tests conduct to different results depending on the retained benchmark requires additional refinement of the findings reported in the literature, which are probably too optimistic regarding the CEECs' real interest rate convergence with the European rate. As shown in previous researches, the interest rate differential between the CEECs and the Euro area seems stationary, confirming the existence of a long term equilibrium relationship, characterizing integrated markets. Nevertheless, the interest rate differential between the CEECs does not clearly appear as stationary, leading to questioning the robustness of the standard second generation tests.

This contradiction can be explained in different ways. The rejection of the RIP can be caused by the existence of a nonstationary risk premium (Chung and Crowder, 2004). In fact, the risk premium $\gamma_t$ reflected in the generalized expression of the UIP (Equation (1)), must be stationary so that the RIRD is stationary in its turn. Or, any risk premium is based on a unit price of the risk, settled on international capital markets when they are integrated, or settled on national or regional markets when they are segmented, and which depends in particular on the investors' riskaversion, which can be reasonably assumed as stationary, and on an amount of risksrelated to macroeconomic factors such as inflation, for which the stationarity assumption has been much more questionable over the past fifteen years, in the case of the CEECs. The disinflation process initiated in these countries, supporting their European integration, could be at the origin of the nonstationarity of the risk premium. Another explanation for the RIP rejection in the case of the CEECs lies in the rejection of the RPPP in the long run, because the $\varepsilon_t$ deviations from the RPPP (Equation (4)) are divergent in the long run.